# Emergent Biological Principles and the Computational Properties of the Universe

P C W Davies
Australian Centre for Astrobiology
Macquarie University, New South Wales, Australia 2109
pdavies@els.mq.edu.au

## ABSTRACT

**The claim that life is an emergent phenomenon exhibiting novel properties and principles is often criticized for being in conflict with causal closure at the microscopic level. I argue that advances in cosmological theory suggesting an upper bound on the information processing capacity of the universe may resolve this conflict for systems exceeding a certain threshold of complexity. A numerical estimate of the threshold places it at the level of a small protein. The calculation supports the contention that "life is an emergent phenomenon."**

**Keywords: Cosmology, computing, complexity, biological organizing principles**

## PHILOSOPHICAL BACKGROUND

The term emergence is used to describe the appearance of new properties that arise when a system exceeds a certain level of size or complexity, properties that are absent from the constituents of the system. It is a concept often summed up by the phrase that "the whole is greater than the sum of its parts," and it is a key concept in the burgeoning field of complexity science. Life is often cited as a classic example of an emergent phenomenon: no atoms of my body are living, yet I am living (see, for example, Morowitz, 2002). Biological organisms depend on the processes of their constituent parts, yet they nevertheless exhibit a degree of autonomy from their parts (see, for example, Kauffman, 2000). How can this be? These seem to be contradictory properties.

The problem of emergence – either explaining it, or explaining it away – has generated a considerable literature (for introductory reviews, see Holland, 1998; Johnson, 2001). Philosophers like to distinguish between weak and strong forms of emergence (Bedau, 1997, 2002). Roughly speaking, a weakly emergent system is one in which the causal dynamics of the whole is completely determined by the causal dynamics of its parts (together with information about boundary conditions and the intrusion of any external disturbances), but for which the complete and detailed behaviour could not be predicted without effectively performing a one-to-one simulation. So a weakly emergent system is one that has no explanatory "short cuts" or abbreviated descriptions, and is therefore algorithmically incompressible in the Kolmogorov-Chaitin sense (Chaitin, 1987). The fastest simulator of the system is the system's own dynamics. Few scientists would quibble with the claim that life is at the very least a weakly emergent phenomenon.

A strongly emergent system is one in which higher levels of complexity possess genuine causal powers that are absent from the constituent parts. That is, wholes may exhibit properties and principles that cannot be reduced, even in principle, to the cumulative effect of the properties and laws of the components. A corollary of strong emergence is the presumption of "downward causation" (Campbell, 1974, Bedau, 2002) in which wholes have causal efficacy over parts. Strong emergence is a much more contentious topic, although there have been many distinguished physicists prepared to argue for some form of it, including founders of quantum mechanics such as Bohr (1933), Schrödinger (1944) and Wigner (1961), as well as some contemporary physicists (e.g. Laughlin et. al., 2000). These strong emergentists are not vitalists: they do not claim that "biotonic laws" (to use Wigner's expression) or "mesoscopic organizing principles" (to use Laughlin's expression) *over-ride* the underlying laws of physics, merely that they *complement* them. Emergent laws of biology may be consistent with, but not reducible to, the normal laws of physics operating at the microscopic level.

Strong emergence has also been advocated by some distinguished philosophers, such as Popper (Popper and Eccles, 1977) and C.D. Broad, who supported its application to biology (Broad, 1925):

> "I know no reason whatever why new and theoretically unpredictable modes of behaviour should not appear at certain levels of complexity, or why they must be explicable in terms of elementary properties and laws of composition which have manifested themselves in less complex wholes… And we have no right to suppose that the laws which we have discovered by studying non-living complexes can be carried over without modification to the very different case of living complexes. It may be that the only way to discover the laws according to which the behaviour of the separate constituents combines to produce the behaviour of the whole in a living body is to study living bodies as such."

In spite of this support, strong emergence is often dismissed as inconsistent with the causal properties of the micro-world. The normal laws of physics operating at the micro-level are supposed to be sufficient to completely determine the behaviour of the system, and so leave "no room at the bottom" for additional biotonic laws or organizing principles to exercise downward causation on the parts. To be sure, it may prove completely impracticable to account for the macroscopic behaviour of, say, a living cell or even a protein by applying the laws of physics at the level of its constituent atoms, but critics of strong emergence retort that, *in principle*, such an account could be given.

An operational distinction between weak and strong emergence is whether or not such an "in principle" microscopic account is possible, notwithstanding its overwhelming mathematical complexity. Sometimes this is cast in the language of Laplace's demon. Laplace, it will be recalled, pointed out that the states of a closed deterministic system, such as a finite collection of particles subject to the laws of Newtonian mechanics, are completely fixed once the initial conditions are specified. Specifically, he wrote (Laplace, 1825)

"We may regard the present state of the universe as the effect of its past and the cause of its future. An intellect which at any given moment knew all of the forces that animate nature and the mutual positions of the beings that compose it, if this intellect were vast enough to submit the data to analysis, could condense into a single formula the movement of the greatest bodies of the universe and that of the lightest atom; for such an intellect nothing could be uncertain and the future just like the past would be present before its eyes."

A strongly emergent system would be one that resisted prediction even by Laplace's demon, quaintly re-named "an archangel" by Broad (1925):

"If the emergent theory of chemical compounds be true, a mathematical archangel, gifted with the further power of perceiving the microscopic structure of atoms as easily as we can perceive hay-stacks, could no more predict the behaviour of silver or of chlorine or the properties of silver-chloride without having observed samples of those substances than we can at present. And he could no more deduce the rest of the properties of a chemical element or compound from a selection of its properties than we can."

And, by extension, the archangel/demon could not deduce the properties of living systems from their molecular components alone. It is this form of *predictive emergentism* that is affected by recent developments in physics and cosmology.

## MATHEMATICS, COMPUTATION AND THE NATURE OF PHYSICAL LAW

Physics is predicated on the assumption that the fundamental laws of the universe are mathematical in nature. Therefore the description, or prediction, of the behaviour of a physical system is implemented by mathematical operations. These operations are necessarily idealizations; for example, the use of differential equations assumes the continuity of spacetime on arbitrarily small scales, the frequent appearance of numbers like $\pi$ implies that their numerical values may be computed to arbitrary precision by an infinite sequence of operations. Many physicists tacitly accept these mathematical idealizations, and treat the laws of physics as implementable in some abstract and perfect Platonic realm. Another school of thought, represented most notably by Wheeler (1984) and Landauer (1967, 1986), stresses that real calculations involve physical objects, such as computers, and take place in the real physical universe, with its specific available resources. In short, information is physical. That being so, it follows that there will be fundamental physical limitations to what may be calculated in the real world, which in turn imposes fundamental limitations on implementing the laws of physics, even in principle. Landauer adopts the position that these limitations are not merely a practical inconvenience, but determine the very nature of physical law (Landauer, 1967):

"The calculative process, just like the measurement process, is subject to some limitations. A sensible theory of physics must respect these limitations, and should not invoke calculative routines that in fact cannot be carried out."

Recall Laplace's description of his calculating demon: "if this intellect were vast enough to submit the data to analysis…" A demon inhabiting an idealized Platonic realm could indeed be "vast enough." But adopting Landauer's view of the nature of physical law, the demon would be obliged to make do with the computational resources of the real universe. Something that could not be calculated, even in principle, within the real universe cannot, according to Landauer, be regarded as a legitimate application of physical law.

The foregoing quibble would not matter for our purposes were it the case that the physical universe possessed infinite computational resources. And there are indeed cosmological models for which there are no limits on the information content and processing power of the universe. However, recent observations favour cosmological models in which there are fundamental upper bounds on both the information content and information processing rate. A Landauer-Laplacian demon associated with such a cosmological model would perforce inherit these limitations, and thus there will exist a fundamental bound on the predictability of complex physical systems, even in principle, if one adopts the Landauer-Wheeler principle of physical law. The next step is to calculate what that bound might be.

## THE FINITE INFORMATION CAPACITY OF THE UNIVERSE

In what follows, I use the word "universe" to refer to a causal region of space. I do not address the issue of whether space as a whole is finite or infinite. Therefore, I assume that the hypothesized link between physical law and the information capacity of the universe refers to information that may be physically processed and accessed in principle from a given location in the universe, e.g. Earth.

Using this definition, the universe is finite in standard big bang cosmological models because of the existence of a particle horizon. Causal horizons occur because of the finite speed of light, and this limits to $\sim (ct)^3$ the volume of space within which information may propagate in the time $t$ since the origin of the universe, assumed to be the origin of time.

Quantum mechanics sets a limit on the speed with which information may be processed by a physical transition, e.g. a spin flip. The maximum rate of elementary operations is $2E/\pi\hbar$, a limit which would be attained by an ideal quantum computer (Margolus and Levitin, 1998).

A third fundamental limit arises because information must either be stored or erased in a finite number of physical degrees of freedom, which imposes a thermodynamic bound (Lloyd, 2000). A convenient way to display this is by combining it with the first two limits in the form of the so-called Bekenstein bound (Bekenstein, 1981):

$kER/\hbar cS \geq 1/2\pi$ (1)

where $k$ is Boltzmann's constant, $R$ is the size of the system (assumed spherical) and $S$ is its entropy. The limit (1) is saturated for the case the case of a black hole, which may be regarded as a perfect information processing (or erasing) system. The information content of the system is related to the entropy $S$ by the Shannon relation:

$$S = k \ln 2. \tag{2}$$

It is straightforward to apply the foregoing limits to a horizon volume within the expanding universe (Lloyd, 2002; see also Penrose, 1979). The total number of bits available within the universe at this epoch is calculated to be $\sim 10^{120}$ if gravitational degrees of freedom are included in addition to all particles of matter. One may also readily calculate the maximum total number of information processing operations that can possibly have taken place since the origin of the universe within an expanding horizon volume. Note that the cosmological scale factor grows like $t^{½}$ initially, whereas the horizon radius grows like $t$. Therefore a horizon volume will have encompassed less particles in the past. Taking this into account, one arrives at an upper bound for the total number of bits of information that have been processed by all the matter in the universe that is also $\sim 10^{120}$ (Lloyd, 2002).

In accordance with the Landauer-Wheeler principle, the enormous but nevertheless finite number $10^{120}$ sets a limit to the fidelity of any in-principle prediction based on deterministic physical law, and hence sets a limit to any constraint of over-determinism that "bottom level" laws of physics might exercise over higher-level, emergent laws. Expressed informally, the existence of an emergent law in a system of sufficient complexity that its behaviour could not be described or predicted by processing $\sim 10^{120}$ bits of information will not come into conflict with any causal closure at the micro-level.

It should be noted that this criterion of permitted emergence is time-dependant: as the universe ages, the horizon volume $\sim (ct)^3$ increases, and with it the effective information capacity of the universe. However, the recent discovery by astronomers of the existence of dark energy implies the presence of a second causal horizon – a de Sitter event horizon, in the simplest case that the dark energy density is constant. The entropy of the de Sitter horizon is given by

$$S_{\text{deS}} = k \times (\text{horizon area})/(\text{Planck area}) = 3kc^5/8G^2\hbar\rho \tag{3}$$

where $\rho$ is the density of dark energy, measured to be about $7 \times 10^{-30}$ gm cm$^{-3}$ $\approx (10^{-3}$ eV$)^4$, $G$ is Newton's gravitational constant, and the Planck area $\equiv G\hbar/c^3 \approx 10^{-66}$ cm$^2$. If the Shannon formula (2) is applied to the de Sitter horizon, it implies an associated information $\sim 10^{122}$ bits. The fact that this number is close to the current information content of the universe is basically the same as the coincidence that, to within a factor of order unity, we find ourselves living at the epoch at which dark energy starts to dominate over matter.

There is strong evidence (e.g. Bousso, 2002; Davies and Davis, 2003) that the de Sitter horizon entropy given by Eq, (3) constitutes an absolute upper bound to the information content of a universe dominated by dark energy of constant energy density (i.e. a cosmological constant). This property has been formally enshrined in the so called *holographic principle* (Susskind, 1995; 't Hooft, 2001; Bekenstein, 2003), according to which the total information content of a simply-connected region of space is captured by the two-dimensional surface that bounds it, after the fashion of a hologram. The maximum information content of a region is given by this surface area divided by the Planck area, which is considered to provide a fundamental finite cell size for space. The de Sitter horizon, which is the end state of cosmological evolution in those models for which dark energy (of constant energy density) eventually dominates, saturates the holographic bound, and so sets an upper limit on the information capacity of the universe throughout its entire evolutionary history. Thus, taking the astronomical observations at face value, it would appear that the universe never contained, and never will contain, more than about $10^{122}$ bits of information, although the number of bits *processed* up to a given epoch will continue to rise with time. Such a rise will not continue indefinitely, however. The holographic bound implies that the universe is a finite state system, and so it will visit and re-visit all possible states (so-called Poincaré cycles) over a stupendous length of time (Goheer et.al., 2003).

It has been suggested (Susskind, 1995) that the holographic principle should be regarded as one of the founding principles of physical science. Combining the holographic principle with the Landauer-Wheeler principle leads inexorably to the causal openness of systems exceeding a certain threshold of complexity.

## HOW COMPLEX MUST A SYSTEM BE TO EXHIBIT GENUINELY EMERGENT BEHAVIOUR?

The numerical results of the previous section may be used to estimate the threshold of complexity beyond which there is no conflict between the causal determinism of the microscopic components and the existence of emergent laws or organizing principles at a higher level.

By way of illustration, consider the oft-cited claim (see, for example, Luisi, 2002) that the enzymatic efficacy of a protein is an emergent property; to wit, that the said efficacy could not be deduced from an application of the laws of physics to a string of amino acids of 20 varieties. Is this correct? A peptide chain of $n$ amino acids can be arranged in $20^n \approx 10^{1.3n}$ different sequences. Each sequence may assume an enormous number of three-dimensional conformations. A crude way to estimate how many is to assume that each amino acid can assume 5 different orientations (Fasman, 1989), so that the number of possible conformations is $5^n \approx 10^{0.7n}$, leading to a total number of possible molecular structures of the order $10^{2n}$. The information processing requirements to explore the chemical properties of each combination are negligible by comparison with the above numbers for even moderate values of $n$, so the predictive limitation is dominated by the combinatorics. Taking the cosmological bound as a lower limit on the "emergence of emergence," we find

$$10^{2n} > 10^{120} \tag{4}$$

or

$$n > 60. \tag{5}$$

This is likely to be an overestimate of the size of the search problem, however, because the conformational information will be algorithmically compressible to some extent, and various approximation schemes may greatly shorten the computational task (e.g. Shirts and Pande, 2001; Garcia and Onuchic, 2003). Stable tertiary structures require conformations that are at least local energy minima – a tiny subset. Real proteins fold much faster than could be explained by their visiting all possible conformations, and so nature has found short cuts to discover stable conformations (although it is unclear whether this is a generic property of peptide chains, or selected for by evolution). A mathematical exploration of the biological efficacy of these molecular shapes is also able to take such short cuts. If the informational burden of the alternative conformations is ignored completely, the bound given by inequality (5) is replaced by $n > 92$. So the simple-minded analysis presented here indicates that the threshold for the onset of emergent properties lies somewhere in the range $60 < n < 92$. Real proteins in fact contain between about 60 and 1000 amino acids, with 100 being typical for a small protein, so the foregoing analysis suggests that protein efficacy may indeed derive, at least in part, from strongly emergent properties.

A similar calculation may be performed for genes. A string of $n$ nucleotides of 4 different bases may be arranged in $4^n \approx 10^{0.6n}$ different combinations, yielding a lower bound for emergent properties of

$$n > 200. \qquad (6)$$

Most genes are somewhat longer than 200 base pairs (typically ~ 1000). For comparison, a small cytoplasmic RNA gene in the human chromosome 7 is a mere 174 base pairs long, but it does not code for a protein.

The fact that these numbers come out so close to the complexity of real biological molecules is striking, given that the limits are derived from fundamental physics and cosmology, and make no reference whatever to biology. The results suggest that the key molecules for life – nucleic acids and proteins – become biologically efficacious at just about the threshold predicted by the Landauer-Wheeler limit corresponding to the onset of emergent behaviour, and that therefore their efficacy may be traced in part to the operation of as-yet-to-be-elucidated biological organizing principles, consistent with, but not reducible to, the laws of physics operating at the micro-level. This analysis therefore supports the contention that life is an emergent phenomenon.

### ACKNOWLEDGEMENTS

I should like to thank John Barrow, Roberto Anitori, Philip Clayton, Neil Rabinowitz and Leonard Susskind for helpful comments.